\documentclass{kapproc} 

\RequirePackage{graphicx}%
\RequirePackage{epsf}%
\input{psfig.sty}

\upperandlowercase
\let\footnote\savefootnote
\let\footnotetext\savefootnotetext
\let\footnoterule\savefootnoterule
\kluwerbib 

\begin{document}

\articletitle{X-rays and young clusters}

\articlesubtitle{Membership, IMFs and distances}

\chaptitlerunninghead{X-rays, IMFs and distances}

\author{Feigelson E.~D. and Getman K.~V.}

\affil{Department of Astronomy \& Astrophysics,
Pennsylvania State University, 525 Davey Laboratory,
University Park, PA 16802, USA}

\email{edf@astro.psu.edu, gkosta@astro.psu.edu}

\begin{abstract}
 
Sensitive imaging X-ray observations of young stellar clusters
(YSCs, ages $\leq 10$ Myr) are valuable tools for the acquisition
of an unbiased census of cluster members needed for Initial Mass
Function (IMF) studies.  Several dozen YSCs, both nearby and across
the Galactic disk, have been observed with the Chandra and
XMM-Newton satellites, detecting $>10,000$ low-mass cluster
members.  Many of these samples should be nearly complete down to
$\simeq 1$ M$_\odot$.  An important additional benefit is that the
YSC X-ray luminosity function appears to be universal with a
lognormal shape, providing a new standard candle for measurement of
YSC distances.

\end{abstract}

\section{Measuring IMFs in young stellar clusters}

Measuring the stellar Initial Mass Function (IMF) requires
complete and unbiased samples which are often sought in young
stellar clusters (YSCs) that have not undergone significant
dynamical evolution.  We consider here YSCs with ages $\leq 10$
Myr where most of the lower mass stars are on the convective
Hayashi pre-main sequence (PMS) evolutionary tracks.

A major difficulty in obtaining a reliable census of YSCs is that
the few lying at distances $d < 400$ pc are spatially extended and
not strongly concentrated.  The richest is the Sco-Cen OB
Association with several thousand members, but it subtends several
thousand square degrees in the southern sky.  Others like the
Corona Australis, Perseus or Chamaeleon clouds have dozens of
members and typically subtend $\geq 1$ square degree.  Individual
Taurus-Auriga clouds or Bok globules may be smaller, but their
stellar populations are too poor for IMF studies.

These large angular sizes require that cluster members of nearby
clouds must be efficiently discriminated from the large, often
overwhelming, number of unrelated foreground and background stars.
This discrimination is traditionally achieved using surveys for
stars with H$\alpha$ emission and/or near-infrared photometric
excesses.  These methods efficiently select protostars and
`classical T Tauri' (CTT) stars, but often undersample the large
`weak-lined T Tauri' (WTT) population (Feigelson \& Montmerle
1999).  As the mass-dependency of disk evolution is not
well known, correction to the full PMS population for IMF
study is uncertain.  Analysis of the $K$-band star counts towards
YSCs can give a more complete census but in an indirect fashion
(Lada \& Lada 1995).  It often requires major corrections for
unrelated stars and, as individual members are not individually
identified, the IMF is derived by a model-dependent conversion of
the $K$-band distributions to mass distributions.

IMF studies may be advantageous in more distant YSCs, but here
additional problems present themselves.  The members are fainter
than in closer clusters, so that large telescopes with low-noise
detectors become necessary to detect the lowest mass cluster
members.  As distance increases, it becomes increasingly difficult
to resolve multiple systems.  Both observational and theoretical
studies suggest that most stars form in binary or multiple systems.
A complete census is also hindered by the wide range of extinctions
often exhibited by YSC members.

In light of these constraints, one YSC has unique merits for IMF
studies:  the Orion Nebula Cluster (ONC).  The ONC lies at $d
\simeq 450$ pc, sufficiently close that the best available
near-infrared instrumentation can detect objects down to a few
Jupiter masses.  With $\simeq 2000$ members concentrated in $\simeq
0.1$ square degrees, it is easily studied and rich enough to
populate the IMF up to $\sim 45$ M$_\odot$.  Unlike other YSCs in
the Orion molecular cloud complex, it lies on the near side of the
cloud and has evacuated most of the intervening molecular material,
so line-of-sight absorptions are low.  While the ONC census suffers
some difficulties $-$ mild contamination from unrelated young stars
in the background cloud and from foreground dispersed young
clusters, incomplete enumeration of multiple systems $-$ it
provides a standard for stellar IMF studies that is far above that
achieved for other YSCs.

\section{The role of X-ray surveys}

X-ray emission from PMS was initially predicted from shocks
associated with their stellar wind, but studies with the Einstein
and ROSAT observatories indicated an origin more closely associated
with solar-like magnetic activity (Feigelson \& Montmerle 1999).
The emission is $1-4$ orders of magnitude stronger than typical
main sequence levels, exhibits high-amplitude flares and hotter
plasmas than expected from wind shocks, and is mostly uncorrelated
with the presence or absence of an infrared-excess disk.  X-ray
surveys discovered many WTTs whose population is comparable to the
CTT population even in YSCs associated with active star forming
clouds.  Although useful, surveys with these early satellites were
limited by detector technology:  low-resolution gas proportional
counters or low-efficiency solid state microchannel plates.

\begin{figure}[ht]
\centering
\includegraphics[width=0.8\textwidth]{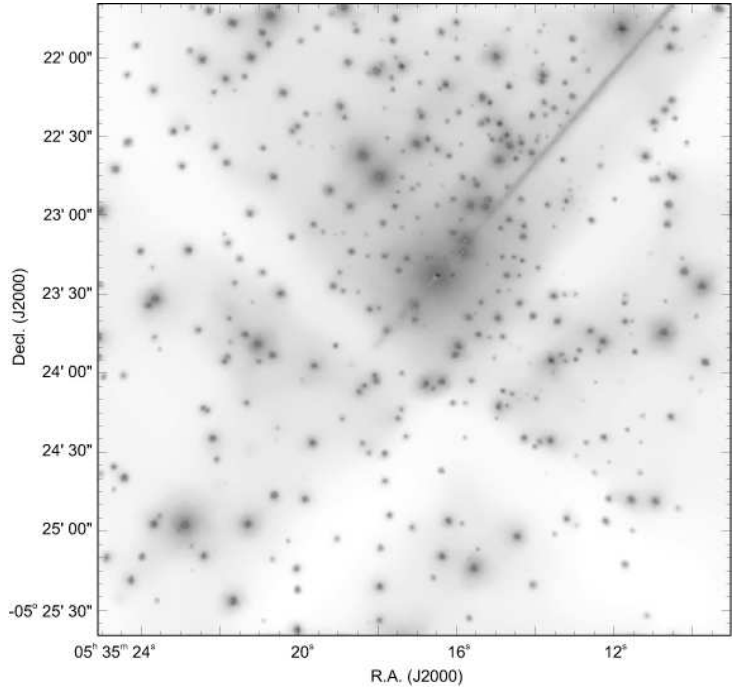}
\caption{The central $4^\prime \times 4^\prime$ region of the
$17^\prime \times 17^\prime$ Chandra ACIS image of the Orion Nebula
Cluster, obtained in a $\simeq 10$-day exposure in early 2003
(Getman et al.\ 2004).  Moderate-resolution spectra over the
$0.5-8$ keV band, giving an independent measure of line-of-sight
absorption, and temporal information on variability are available
for each of the 1616 COUP sources.This dataset is the foundation of
a wide range of studies comprising the Chandra Orion Ultradeep
Project (COUP).  \label{COUPimg.fig}}
\end{figure}

The Chandra X-ray Observatory and XMM-Newton missions are much
better adapted to YSC studies with their high-efficiency low-noise
CCD detectors.  Chandra is particularly useful with its
high-precision mirrors giving arcsecond imaging capability, though
its field of view is limited to 0.08 square degrees.  Together,
these telescopes have imaged several dozen YSC populations across
the Galactic disk.  These include the $\rho$ Ophiuchi, Chamaeleon,
CrA, Mon R2, L1551, L1448, Serpens, Cep B and Sgr B2 clouds;
populations associated with the Orion, Trifid, Rosette and Carina
HII regions; the Cyg OB2 association, $\eta$ Cha cluster, and
various isolated Herbig Ae/Be stars with their companions; star
forming regions NGC 1333, IC 348, M 8, M 16, M 17, NGC 281, NGC
1579, NGC 2078, NGC 2024, NGC 2264, IC 1396, RCW 38, RCW 49, RCW
108, NGC 6334, NGC 6530, NGC 6383, NGC 3603, W 3, W 1, W 49, W 51,
and IRAS 19410+2336; several Galactic Center YSCs; and 30 Dor
in the Large Magellanic Cloud.  The most comprehensive study
underway is the Chandra Orion Ultradeep Project (COUP) based on a
nearly-continuous 10-day pointing towards the ONC and embedded
sources in OMC-1.  An introduction to COUP, and references for the
other regions published through mid-2003, are given in Feigelson
(2003a).  Figure \ref{COUPimg.fig} shows a portion of the COUP
image, and some preliminary COUP results are presented here.

X-ray surveys are subject to selection effects which must be
carefully considered in the effort to achieve a well-defined YSC
census.  Most importantly, PMS X-ray luminosities are strongly
correlated with a tangle of interrelated stellar properties:
bolometric luminosity, mass, stellar surface area and volume.  The
$L_x-L_{bol}$ correlation has been known for many years, but the
linkage with other variables only emerged clearly in pre-COUP
studies of the ONC (Flaccomio et al.\ 2003a; Feigelson et al.\
2003b).  Though quantitative analysis of the scatter in these
relations by the COUP has not yet been completed, we can roughly
say that a Chandra observation of a nearby YSC with limiting
sensitivity $\log L_x \rm{(lim)} \simeq 28.0$ erg/s ($0.5-8$ keV
band) will detect $>90$\% of PMS stars with masses $M > 0.2$
M$_\odot$, and an observation of a more distant YSC with $\log L_x
\rm{(lim)} \simeq 29.5$ erg/s will detect $>90$\% of PMS stars with
$M > 0.8$ M$_\odot$ (Preibisch et al., in preparation).\footnote{A
second second selection effect in PMS X-ray luminosities must be
considered:  the subpopulation of CTT stars tends to be a factor of
$2-3$ weaker the subpopulation of WTT stars leading to, opposite to
traditional methods, a selection bias against accreting stars
(Flaccomio et al.\ 2003b).  But this effect is relatively small
compared to the 4 orders of magnitude range in PMS X-ray
luminosities.}

This result immediately indicates the power of Chandra studies for
improving our knowledge of stellar populations in a wide variety of
YSCs, and thus studying the IMF in a wide variety of conditions.
In most YSCs beyond $d \simeq 1$ kpc, cluster membership is
currently limited to a small number of bright
spectroscopically-confirmed OB stars and lower mass stars with
$K$-band excesses.  Consider, for example, the cluster illuminating
the HII region M 17 and its the surrounding molecular cloud.  Only
a few dozen have optical spectra and while $>$20,000 have $JHK$
measurements, most of these stars are unrelated to the cloud
(Hanson et al.\ 1997; Jiang et al.\  2002).  A Chandra
image with $\log L_x \rm{(lim)} = 29.7$ erg/s shows 877 sources
nearly all of which are cluster members (Getman et al., in
preparation).  Such X-ray selected samples should be $> 90$\%
complete above a well-defined mass limit, usually around $0.5-1.5$
M$_\odot$ for the more distant YSCs under study. 

Analysis of these fields is difficult with hundreds of faint X-ray
stars often embedded in structured diffuse emission from
large-scale OB wind shocks (Townsley et al.\ 2003).  But
sophisticated data analysis methods have been developed (Getman et
al.\ 2004), careful studies are underway and within a few years
$>10,000$ Chandra-discovered YSCs members should be published.
Followup optical/near-infrared photometry and spectroscopy will be
needed to place these stars on the HR diagram to estimate masses
for IMF analysis.  Chandra should thus provide an important boost
to comparative IMF studies for many of the YSC clusters listed
above.

We caution that it is still unclear whether even the most sensitive
X-ray surveys can efficiently detect substellar PMS objects which
will evolve into L- and T-type brown dwarfs.  Two YSC studies with
sensitivities of $\log L_x {\rm(lim)} \simeq 27.0$ erg/s performed
to date give inconsistent results.  A deep Chandra image of the
Chamaeleon I North cloud found all 27 known cloud members including
3 probable substellar objects (Feigelson \& Lawson 2004).  This
suggests the X-ray census is complete and gives an IMF deficient in
stars with $M < 0.3$ M$_\odot$ compared to the ONC or field star
IMFs.  But this may be due to mass segregation favoring higher mass
stars in the small 0.6~pc$^2$ region covered by the Chandra imager,
rather than intrinsically different IMF.  The second study with
$\log L_x {\rm(lim)} \simeq 27.0$ erg/s is the COUP observation of
the ONC.  Here, most of the spectroscopically confirmed brown
dwarfs (Slesnick et al.\ 2004) are not detected, indicating that
X-ray surveys will probably be incomplete in this low mass regime.

\section{A new distance estimator for YSCs}

\begin{figure}[ht]
\centering
\includegraphics[width=0.7\textwidth]{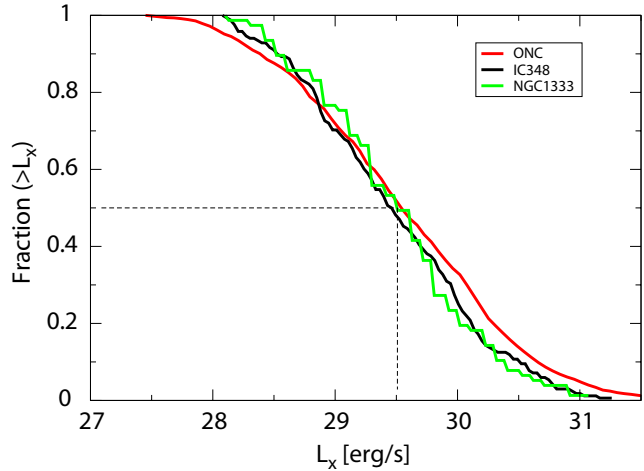}
\caption{The universal cluster X-ray luminosity
function (XLF) found in Chandra YSC studies.  NGC 1333 (77 stars
from Getman et al.\ 2002), IC 348 (168 stars from Preibisch \&
Zinnecker 2002), and the ONC (1508 stars from COUP data truncated
at $\log L_x = 31.5$ erg/s, Getman et al.\ 2004). }
\end{figure}

Techniques for measuring distances to YSCs and their natal
molecular clouds have hardly changed in a half-century.  When a OB
population can be studied spectroscopically and individual stellar
absorptions are found, then main sequence fitting provides a
reliable distance.  But when they are too obscured or unavailable,
then distance estimation methods are varied, unreliable and
inconsistent.  Consider, for example, the YSC surrounding the
Herbig AeBe star LkH$\alpha$ 101, one of the brightest infrared
sources in the sky, which illuminates the nebula NGC 1579.  The
cluster distance has variously been estimated to be 140 pc, $>$800
pc, $\simeq 340$ pc, and $\simeq 700$ pc (Tuthill et al.\ 2002,
Herbig et al.\ 2004).  If it lies at the nearer end of this range,
it is one of the closest YSCs.

The X-ray luminosity function (XLF) of YSCs has two remarkable
empirical characteristics that should render it an effective and
accurate distance estimator for clusters such as this.  First, the
shapes of different YSC XLFs appear to be remarkable similar to
each other, once a richness-linked tail of high luminosity O stars
is omitted (Figure 2).  Second, the shape of this `universal' XLF
strongly resembles a lognormal with mean $<\log L_x> \simeq 29.5$ erg/s
($0.5-8$ keV band) and standard deviation $\sigma (\log L_x) \simeq
0.9$ (Figure 3).

\begin{figure}[ht]
\centering
\includegraphics[width=0.7\textwidth]{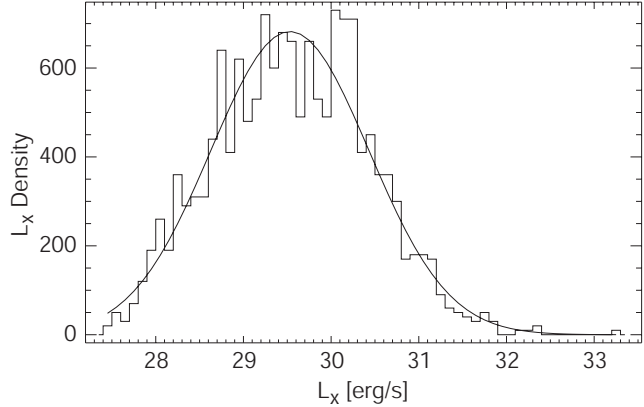}
\caption{The Orion Nebula Cluster (ONC) XLF with lognormal fit given in the
text.  COUP sample of 1528 sources with absorption corrected 0.5-8
keV luminosities (Getman et al.\ 2004).  All X-ray luminosities
are in the $0.5-8$ keV band corrected for absorption measured
from the X-ray spectrum.}
\end{figure}

While we do not have astrophysical explanations for either of these
YSC XLF properties, the shape of the XLF can be roughly understood
as a convolution of the IMF which breaks from the Salpeter powerlaw
below $\simeq 0.5$ M$_\odot$ (and itself is sometimes modeled with
a lognormal curve) and the correlation between $L_x$ and mass.
These two effects result in a steep falloff in the number of
fainter X-ray stars in a YSC.  This was dramatically demonstrated
in the Chandra studies of the ONC where a factor of 10 increase in
limiting sensitivity from $\log L_x {\rm (lim)} \simeq 28.0$ erg/s
in the early observations to $\simeq 27.0$ erg/s in the deep COUP
observation led to only a very small rise in detected lightly
absorbed ONC stars.

The peak of the XLF at $<\log L_x> \simeq 29.5$ erg/s can be used
as a standard candle for distance estimation, much as the lognormal
distribution of globular cluster optical luminosities is used as an
extragalactic distance estimator (Harris 1991).  One counts cluster
members in flux units (this is the X-ray $\log N - \log S$
curve), matches the observed $\log N - \log S$ to the universal XLF, 
and reads off the distance from the offset.  The table below gives a
simple example for a hypothetical cluster at various assumed
distances observed with Chandra for 100 ks.  The tabulated values
are the fraction of sources seen in broad count rate bins assuming
typical PMS X-ray spectra and negligible absorption.

\begin{table}[ht]
\caption{Simulated X-ray source counts for young stellar clusters}
\begin{tabular*}{\textwidth}{@{\extracolsep{\fill}}crrrrr}
\sphline
$\log L_x \rm{(lim)}$ (erg/s) & 28.0 & 28.5 & 29.0 & 29.5 & 30.0\cr 
Distance (pc) & 400 & 710 & 1270 & 2250 & 4000\cr
Flux bin (cts) & \multicolumn{5}{c}{Percent of sources} \cr \sphline
$5 - 50$       & 22 & 46 & 60 & 76 & 86 \cr 
$50 - 500$     & 46 & 42 & 34 & 22 & 14 \cr 
$500 - 5000$   & 28 & 12 &  6 &  2 &  0 \cr 
$5000 - 50000$ &  4 &  0 &  0 &  0 &  0 \cr \sphline
\end{tabular*}
\end{table}

The table shows that the observed $\log N - \log S$ shape will
differ dramatically for YSCs at different distances.  For close
distances or unusually long exposures (as in COUP), most cluster
members are quite bright and few faint ones are seen.  For far
distances or short exposures, most cluster members are near the
detection limit which is $3-10$ photons for typical Chandra fields.
The principal challenges are corrections for absorption, which can
be derived (except for the faintest sources) for individual sources
from the X-ray spectrum, and the elimination of extragalactic
background sources, which can be achieved by the absence of a
stellar counterpart in sensitive $K$-band images.  Accurate
distances and error analysis would not be performed in broad flux
bins as in the table above, but would be based on unbinned sources
fluxes using a maximum likelihood method as described by Hanes \&
Whittaker (1987) and Cohen (1991).

{\it Acknowledgements:}  We thank Thomas Preibisch (MPIfR) for use
of unpublished COUP results.  This work was supported by NASA
contract NAS8-38252 (Garmire, PI) and COUP grant SAO GO3-4009A
(Feigelson, PI).

\end{document}